\documentclass[onecolumn,showpacs,preprintnumbers, sort&compress, floatfix]{revtex4}
\usepackage{amsmath}
\usepackage{amsfonts}
\usepackage{amssymb}
\usepackage{graphicx}
\usepackage{bm}
\def\be{\begin{equation}}
\def\ee{\end{equation}}
\def\bi{\bibitem}

\begin{document}
\title{History of cosmic evolution with modified Gauss-Bonnet-dilatonic coupled term.}
\author{Subhra Debnath}
 \email{subhra_dbnth@yahoo.com}
 \affiliation {Dept.of Physics, Jangipur College, Murshidabad, India - 742213}
\author{Soumendra Nath Ruz}
 \email{ruzfromju@gmail.com}
 \affiliation {Dept.of Physics, Ramananda Centenary College, Purulia, India - 723151}
\author{Ranajit Mandal}
 \email{ranajitmandalphys@gmail.com}
 \affiliation {Dept.of Physics, University of Kalyani, Nadia, India - 741235}
\author{Abhik Kumar Sanyal}
 \email{sanyal_ak@yahoo.com}
 \affiliation {Dept.of Physics, Jangipur College, Murshidabad, India - 742213}

\begin{abstract}
\noindent Gauss-Bonnet-dilatonic coupling in four dimension plays an important role to explain late time cosmic evolution. However, this term is an outcome of low energy string effective action and thus ought to be important in the early universe too. Unfortunately, phase-space formulation of such a theory does not exist in the literature due to branching. We therefore consider a modified theory of gravity, which contains a nonminimally coupled scalar-tensor sector in addition to higher order scalar curvature invariant term with Gauss-Bonnet-dilatonic coupling. Such an action unifies early inflation with late-time cosmic acceleration. Quantum version of the theory is also well-behaved.
\end{abstract}


\maketitle

\section{Introduction}

A smooth Luminosity-distance versus redshift curve reveals that distant Supernovae appear dimmer than usual \cite{Riess}, \cite{Pulmutter}. This issue may be explained by modifying the energy-momentum tensor appearing on the right hand side of Einstein's equation, giving rise to the so-called dark energy. Almost all the solutions to Einstein's equations result in accelerated expansion of the universe, if some form of dark energy is invoked. As a result, Supernovae data, dark energy and late-time cosmic acceleration became synonym. Nevertheless, Friedmann-like matter dominated era ($a(t) \propto t^{2\over 3}$, $a(t)$ being the scale factor), should be followed by such accelerated expansion, which is a recent phenomena, otherwise, it would tell upon the structures we observe. Further, the solution should also match other experimental data, e.g. matter-radiation equality at redshift, $z \approx 3200$ \cite{Komatsu}, decoupling at $z \approx 1080$ \cite{Galli} and anisotropy of CMBR, released from WMAP data \cite{Spergel}. These experimental data rule out some of the dark energy models. However, the problem with the remaining models is two fold: firstly, the models are indistinguishable from each other, and second, complicated models with two of more scalar fields and sometimes with reverse sign of kinetic energy is required to exhibit crossing of phantom divide line. In this sense, a better option is to modify the left hand side of Einstein's equation by incorporating higher order curvature invariant terms in the Einstein-Hilbert action. Such a theory is dubbed as modified theory of gravity. A successful modified theory of gravity was first proposed by Nojiri and Odintsov \cite{Odintsov}, in the form
\be \label{A1} A_1 = \int d^4x \sqrt{-g}F(R) = \int d^4x \sqrt{-g}\left[{R\over 16\pi G} + \beta R^2 + \gamma R^{-1}\right]\ee
where, $R^2$ term dominates at the very early universe, leading to inflation, $R$ dominates in the middle, so that Nucleosynthesis, CMBR, structure formation etc. remain unaltered from standard model, and $R^{-1}$ dominates at the late-stage of cosmic evolution, leading to late time acceleration of the universe. A scalar mode called ``Scalaron" appears due to scalar-tensor equivalence of higher order theory. The mass of the scalar field may be adjusted, by suitably fine tuning $\beta$, so that the model passes solar test. Thus, a model which appears to reconcile early inflation with late-time cosmic acceleration satisfying all the presently available experimental data, bypassing dark energy issue, is at hand. The only problem is, $R^{-1}$ term is not recognized at all from any physical argument. The same has also been attempted latter by Modak, Sarkar and Sanyal \cite{Modak} successfully, with an action
\be \label{A11} A_1 = \int d^4x \sqrt{-g}F(R) = \int d^4x \sqrt{-g}\left[{R\over 16\pi G} + \beta R^2 + \gamma R^{3\over 2}\right]\ee
which explains reionization, in addition. Although $R^{3\over 2}$ is an artefact of Noether symmetry, however none of these terms ($R^{-1}$ or $R^{3\over 2}$) is generated by one-loop quantum gravitational corrections. Therefore, even slightest presence of these terms in the early universe shatters all attempts to obtain a renormalized theory of gravity.\\

Attempt to modify gravity was initiated almost a century back by Weyl \cite{Weyl}, soon after the advent of general theory of relativity (GTR). Latter, it was realized that the inevitable gravitational collapse is due to application of GTR beyond its domain of validity. However, attempt to find a quantum counterpart at the Planck's scale or beyond, revealed that GTR suffers from ultraviolet divergence and so it is not renormizable. A renormalized theory of gravity by incorporating curvature squared terms in Einstein-Hilbert action in the form
\be\label{A2}A_1 = \int d^4x \sqrt{-g}\left[{R\over 16\pi G} + \beta R^2 + \gamma R_{\mu\nu}R^{\mu\nu}\right]\ee
was presented in the late twentieth century \cite{Stelle}. Unfortunately, this came at a very high price, since fourth order derivatives lead to ghosts in the perturbation series about the linearized theory. Attempt to construct a second order theory out of higher order curvature invariant terms lead to a particular combination viz. the Gauss-Bonnet combination, $\mathcal{G} = R^2 - 4 \gamma R_{\mu\nu}R^{\mu\nu} + \gamma R_{\mu\nu\delta\gamma}R^{\mu\nu\delta\gamma}$. Although one may consider the (Wald) entropy effect of  Gauss-Bonnet (GB) gravity in four dimensions \cite{10}, it is notable that the variation of GB Lagrangian is a total derivative in four dimension, and therefore it does not contribute to the four dimensional field equations, as well as to the black hole solutions \cite{11}. Thus, in order to study the contributions of GB term, solutions to Lanczos-Lovelock gravity \cite{LL}, which is realized in five and even higher dimensions, is studied.\\

There exists a variety of important features of GB Gravity. Its effects on the speed of graviton propagation together with the appearance of potentially super-luminal modes have been investigated \cite{Ycb} and the problem with its unusual causal structure has also been resolved \cite{Kizumi}. It was further applied to investigate possible resolution of the initial singularity and graceful exit problem \cite{Kanti}. Stability criteria has also been studied \cite{stab}. In addition, the influences of GB gravity have been investigated as regards various physical phenomena such as superconductors \cite{4}, hydrodynamics \cite{5}, LHC black holes \cite{6}, dark matter \cite{7}, dark energy \cite{8} and shear viscosity \cite{9}. \\

If one therefore wants to restrict to four dimensions with Gauss-Bonnet term, it is possible with Gauss-Bonnet dilatonic coupled term, which arises naturally as the leading order of the $\alpha'$ expansion of heterotic superstring theory, where, $\alpha'$ is the inverse string tension \cite{GB}. Such a term works fairly well in four dimension. e.g. it admits Black-Hole solutions \cite{black} and gives rise to cosmic inflation at the early epoch \cite{Neupane}. Late time accelerated expansion of the universe is particularly a very special feature of such action \cite{late, Neupane, Pierre}. Moreover, important issues like - late time dominance of dark energy after a scaling matter era and thus alleviating the coincidence problem crossing the phantom divide line, and compatibility with the observed spectrum of cosmic background radiation have also been addressed recently \cite{gba9, san}. It also gives fruitful results in Noether symmetry study as well \cite{aks}. In a nutshell, gravitational action containing Einstein-Gauss-Bonnet-dilatonic coupling has been able to explain the cosmological puzzle, successfully. \\

Here, we therefore aim to expatiate the effect of Einstein-Gauss-Bonnet-dilatonic coupled action in the very early universe, following canonical quantization. This primarily requires canonical formulation of the theory. However, while performing canonical analysis of Lanczos-Lovelock action \cite{LL} under $3+1$ decomposition, Deser and Franklin noticed that the presence of cubic kinetic terms and quadratic constraints, make the theory intrinsically nonlinear \cite{deser}. The pathology associated with Lovelock AdS Black Branes and AdS/CFT was also discussed by Takahashi and Soda \cite{taka}. So one can not perform standard Hamiltonian formulation of such an action following conventional Legendre transformation. As a result, diffeomorphic invariance is not manifest and standard canonical formulation of the theory is obscure. Such a situation arises because the Lagrangian is quartic in velocities, and therefore the expression for velocities are multivalued functions of momentum, resulting in the so called multiply branched Hamiltonian with cusps. This makes classical solution unpredictable, as at any instant of time one can jump from one branch of the Hamiltonian to the other, as equation of motion allows such jumps. It is important to mention that in principle, the particle always remains in one branch of the Hamiltonian and the presence of cusps restricts the domain of the variables of the problem. To make things more apparent, let us take help of the following toy model following \cite{Teit, Che}
\be\label{ta} A = \int \Big[{1\over 4}\dot q^4 - {1\over 2}\alpha\dot q^2\Big]dt,\ee
for which
\be \label{te}{dp(\dot q)\over dt} = 0;\;\;\;p(\dot q) = p_0 = \dot q^3 - \alpha \dot q.\ee
It is now required to solve $\dot q$ as a function of integration constant $p_0$, which is possible uniquely for $\alpha < 0$ as depicted in figure 1. However, for $\alpha > 0$, the inverse image of the velocity $\dot q$ is single valued if initial momentum ($p_0$) lies beyond the closed interval $[p_1, p_2]$; doubly valued, if initial momenta is at one of the critical values $p_1$ or $p_2$, and finally within the interval, it takes three values, as depicted in figure 2. Therefore within the interval, for a particular value of $p$, one doesn't know which initial value of $\dot q$ it belongs to, and the equations of motion allows instantaneous switching (jumps) from one $\dot q$˙value to another, since these instantaneous jumps leave $p$ unchanged and satisfy equation of motion. Further, since each given $p$ corresponds to one or three values of $\dot q$, the Hamiltonian
\be \label{th}H = {3\over 4}\dot q^4 - {\alpha\over 2} \dot q^2\ee
is also mulltivalued with cusps, as shown in figure 3. At any instant of time one does not know which ``branch" of the Hamiltonian to use, thus one may propagate for a while with one choice of the Hamiltonian, then switch to another and so on. Since the switching may be done after arbitrarily small time intervals, one may visualize the classical motion as a succession of zigzags which happen in an unpredictable manner. Thus, the behavior of the system described by the action \eqref{ta} remains unpredictable for a range of initial data of non-vanishing extent. Further, the momentum does not provide a complete set of commuting observable resulting in non-unitary time evolution of quantum states.
\begin{figure}
\begin{center}
\includegraphics[height = 2.0 in, width = 3.0 in]{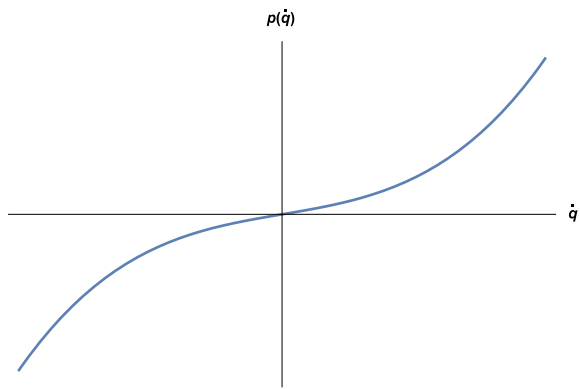}
\end{center}
\caption{For $\alpha < 0$, Legendre map from $\dot q$ to $p$ shows one to one correspondence between the two.}
\end{figure}
\begin{figure}
\begin{center}
\includegraphics[height = 2.0 in, width = 3.0 in]{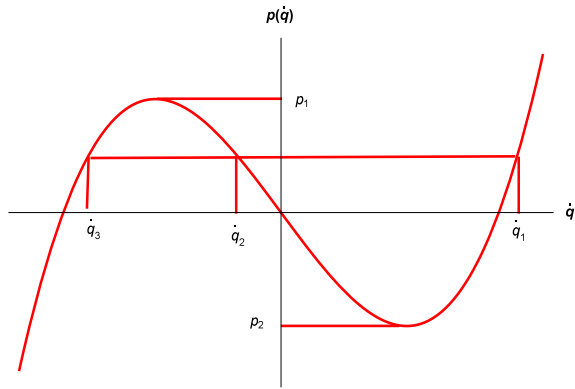}
\end{center}
\caption{For $\alpha > 0$, Legendre map from $\dot q$ to $p$ shows multivaluedness of inverse images $\dot q$.}
\end{figure}
\begin{figure}
\begin{center}
\includegraphics[height = 2.5 in, width = 3.0 in]{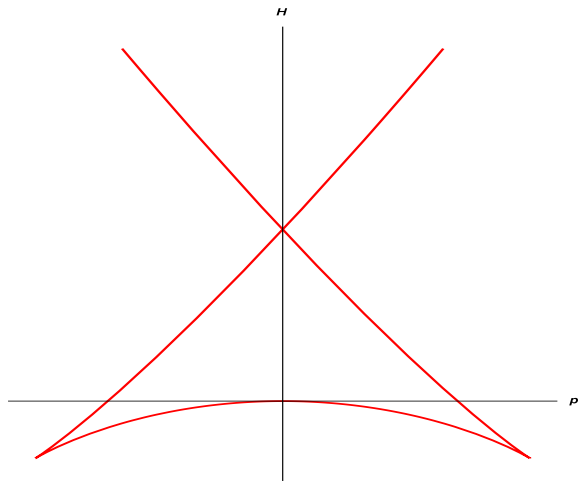}
\end{center}
\caption{Phase-space portrait depicts multivaluedness of the Hamiltonian $H$ with cusps.}
\end{figure}

Although, Lanczos-Lovelock gravity \cite{LL} shows unitary time evolution of quantum  states, when expanded perturbatively about the flat Minkowski background; non-perturbatively, the situation is miserable. In the following section-II, we show that Einstein-Gauss-Bonnet-dilatonic action suffers from the same disease. Presently, there is no standard technique to handle this issue. Two recent techniques in this fields are Legendre-Fenchel transformation \cite{Che} and Generalized Legendre Transformation \cite{Abra}. However, it has been shown that for the same system, the Hamiltonian obtained following the two techniques are not related through canonical transformation \cite{aks1}. Thus, one does not know which one is the correct Hamiltonian. Nevertheless, one can bypass the issue of branching arising out of higher degree, by incorporating higher order terms \cite{aks1a}. This has been addressed recently, to alleviate the problem of branching in Lanczos-Lovelock gravity \cite{LL}, by modifying the action, associating a scalar curvature invariant term  \cite{aks1}. Here in section-III, we follow the same procedure in homogeneous and isotropic minisuperspace model, to cast the modified action
\be\label{action} A=\int\sqrt{-g}\;d^4x\left[\frac{R}{16\pi G} + \xi(\phi)\left(\mathcal{G}+ \beta R^2 \right) - \frac{1}{2} \phi_{,\mu}\phi^{,\mu}-V(\phi)\right] \ee
in canonical ADM form \cite{ADM}. Quantum description of the theory leads to Schr\"odinger-like equation, where, the proper volume acts as an internal time parameter. The effective Hamiltonian operator has been found to be hermitian following appropriate ordering prescription, for which probabilistic interpretation is straight forward. Semi-classical approximation depicts oscillatory behaviour of the wave function about a classical inflationary solution. It therefore appears that action \eqref{action} is a better option to demonstrate evolutionary history of the universe.

\section{Problem in Canonization of Gauss-Bonnet dilatonic coupling}
As already mentioned, Gauss-Bonnet-dilatonic coupled term appears as the leading order in low energy effective Heterotic super-string theory. Although, it can explain late-time cosmological evolution and fits other cosmological data when associated with Einstein-Hilbert term, it contributes in the early universe as well. Therefore, it is important to study the quantum cosmological aspect of such a term. This requires Hamiltonian formulation of the theory. In this section we aim to show that this is a non-trivial issue even in the homogeneous and isotropic minisuperspace model. The action we start with is

\be\label{A1} A_1=\int\sqrt{-g}\;d^4x\left[\frac{R}{16\pi G}+\xi(\phi)\mathcal{G} - \frac{1}{2} \phi_{,\mu}\phi^{,\mu}-V(\phi)\right] + \Sigma_R + \xi(\phi)\Sigma_\mathcal{G},\ee
where
\begin{subequations}\begin{align}
\Sigma_R &= \frac{1}{8\pi G}\oint_{\partial\mathcal{V}}K \sqrt hd^3x \\
\Sigma_\mathcal{G} &= 4\oint_{\partial\mathcal{V}} \left( 2G_{ij}K^{ij} + \frac{\mathcal{K}}{3}\right)\sqrt hd^3x
\end{align}\end{subequations}
are the boundary terms corresponding to $R$ and $\mathcal{G}$ respectively. Here $\mathcal{G}$, is coupled with $\xi(\phi)$, and $V(\phi)$ is the dilatonic potential. The symbol $\mathcal{K}$ stands for $\mathcal{K}=\left( K^3 - 3K K^{ij}K_{ij} + 2K^{ij}K_{ik}K^k_j \right)$  where, $K$ is the trace of the extrinsic curvature tensor $K_{ij}$, and $G_{ij}$ is the Einstein tensor built out of the induced metric $h_{ij}$ on the boundary. In the homogeneous and isotropic Robertson-Walker metric, viz.,

\be\label{rw} ds^2 = - N(t)^2 dt^2 + a^2(t) \left[\frac{dr^2}{1-kr^2} + r^2 (d\theta^2 + sin^2 \theta d\phi^2)\right],\ee
the expressions for $R$ and $\mathcal{G}$ are
\begin{subequations}\begin{align}\label{rg}
R &= \frac{6}{N^2}\left(\frac{\ddot a}{a}+\frac{\dot a^2}{a^2}+N^2\frac{k}{a^2}-\frac{\dot N\dot a}{N a}\right) \\
\mathcal{G} &= \frac{24}{N^3 a^3}\left(N\ddot a - \dot N \dot a\right)\left(\frac{\dot a^2}{N^2} + k\right)
\end{align}\end{subequations}
Let us first express the action (\ref{A1}) in terms of the basic variable, $h_{ij} = a^2 = z$. This choice helps not only to take care of the boundary terms, but also allows to cast the action in ADM form as well \cite{aks2}. In terms of the new variable, the action (\ref{A1}) reads,
\be\begin{split}\label{6}
A_1 &= \int\Big[\frac {3\sqrt z}{16 \pi G}\left(\frac{\ddot z}{ N}- \frac{\dot N \dot z}{N^2} + 2k N \right) + \frac {3 \xi(\phi)}{N \sqrt z}\left(\frac{{\dot z}^2 \ddot z}{N^2 z} + 4 k \ddot z - \frac{{\dot z}^4}{ 2 N^2 z^2} - \frac{\dot N {\dot z}^3}{N^3 z} - \frac{2 k {\dot z}^2}{z}- \frac{4 k \dot N \dot z}{N}\right) \\
& + z^{\frac{3}{2}}\left(\frac {{\dot {\phi}}^2}{2N} - VN\right)\Big]dt + \Sigma_R +\xi(\phi)\Sigma_\mathcal{G}.
\end{split}\ee
Here
\begin{subequations}\begin{align}
\Sigma_R &= - \frac{3 \sqrt z ~ \dot z}{16 \pi G N} \\
\Sigma_\mathcal{G} &= -\frac {\dot z}{N \sqrt z}\Big(\frac{{\dot z}^2}{N^2 z} + 12 k\Big).
\end{align}\end{subequations}
Under integration by parts, the total derivative terms get cancelled with the boundary term and the action (\ref{6}) takes the form

\be A_1 = \int\left[\frac {1}{16 \pi G}\left( - \frac{3 {\dot z}^2}{2 N \sqrt z} + 6 k N \sqrt z \right) - \frac {\xi^{'} \dot z \dot {\phi}}{N \sqrt z}\left(\frac{{\dot z}^2}{N^2 z} + 12 k\right) + z^{\frac{3}{2}}\left(\frac {1}{2N} {\dot {\phi}}^2 - VN\right)\right]dt.\ee
Canonical momenta are
\begin{subequations} \begin{align}
& p_z = - \frac{3 {\dot z}}{16 \pi G N \sqrt z}- \frac {3 \xi^{'} \dot {\phi}}{N \sqrt z}\left(\frac{{\dot z}^2}{N^2 z} + 4 k\right) \\
& p_\phi = - \frac {\xi^{'} \dot z}{N \sqrt z}\left(\frac{{\dot z}^2}{N^2 z} + 12 k\right) + \frac{z^{\frac{3}{2}} \dot {\phi}}{N}  \\
& p_N = 0
\end{align} \end{subequations}
The $N$ variation equation,
\be
-\frac {3}{16 \pi G}\left(\frac{{\dot z}^2}{2 N^2 \sqrt z} + 2 k \sqrt z \right) - \frac {3 \xi^{'} \dot z \dot {\phi}}{N^2 \sqrt z}\left(\frac{{\dot z}^2}{N^2 z} + 4 k\right) + z^{\frac{3}{2}}\left(\frac {1}{2N^2} {\dot {\phi}}^2 + V\right) = 0
\ee
leads to the Hamiltonian constraint equation, which is essentially the $\left(^0_0\right)$ equation of Einstein, viz.
\be
H_c = N \left[ -\frac {3}{16 \pi G}\left( \frac{{\dot z}^2}{2 N^2 \sqrt z} + 2 k \sqrt z \right) - \frac {3 \xi^{'} \dot z \dot {\phi}}{N^2 \sqrt z}\left(\frac{{\dot z}^2}{N^2 z} + 4 k\right) + z^{\frac{3}{2}}\left(\frac {1}{2N^2} {\dot {\phi}}^2 + V\right)\right] = 0.
\ee
The definitions of momenta imply that they are up-to fifth degree equations in $\dot z$ and so are not invertible in general. Therefore, action (\ref{6}) suffers from the problem of branching as discussed in the introduction. Hence, it is impossible to cast the Hamiltonian in canonical form and despite diffeomorphic invariance of the theory, it is not possible to express the Hamiltonian as $H_c = N\mathcal{H}$.

\section{Modified dilatonic coupled Gauss-Bonnet action}

A host of techniques are presently available in the literature to get rid of the above awesome situation, as already cited. Although these techniques alleviate the problem of branching, no two phase-space Hamiltonian are related through canonical transformation, and therefore no one knows which is the correct description of the theory under consideration. However, the situation is considerably improved bypassing the associated problem, by incorporating additional higher order curvature invariant term in the action, as already noticed earlier \cite{aks1, aks1a}. Under appropriate field redefinition followed by conformal transformation, it is always possible reduce the string frame action (to the first order in $\alpha'$) to the form in Einstein's frame given in \eqref{action}. Such a field redefinition is usually performed to get rid of higher-order derivatives from the field equations. Here our purpose is to demonstrate the fact that scalar curvature squared term can alleviate the problem of branching. Therefore, we consider an additional $R^2$ term in the form given in \eqref{A1}. Including appropriate boundary terms such an action reads,

\be\label{A2}A_2=\int\sqrt{-g}\;d^4x\left[\frac{R}{16\pi G} +  \xi(\phi)\left(\beta R^2 +\mathcal{G}\right) - \frac{1}{2} \phi_{,\mu}\phi^{,\mu}-V(\phi)\right] + \Sigma_R + \xi(\phi)[\Sigma_\mathcal{G} +  \beta(\Sigma_{R^2_1} + \Sigma_{R^2_2})],\ee
where
\be \label{100}\Sigma_{R^2} = \oint_{\partial\mathcal{V}} {^4R} K\sqrt h d^3x = \Sigma_{R^2_1} + \Sigma_{R^2_2} = \oint_{\partial\mathcal{V}} {^3R} K \sqrt h d^3x + \oint_{\partial\mathcal{V}} (^4R - {^3R}) K\sqrt h d^3x, \ee
${^3R}$ being the Ricci scalar in 3-space . In the Robertson-Walker minisuperspace (\ref{rw}) under consideration,
\be \Sigma_{R^2_1} = -36 k \frac{\dot z}{N \sqrt z} \ \ \text{and} \ \ \Sigma_{R^2_2} = -18\frac{\dot z}{N^3 \sqrt z}\left(\ddot z  - \frac{\dot N \dot z}{N}\right) \ee
The reason for splitting the boundary term in two parts, has been discussed in details earlier \cite{aks2, GRG}. Now under integration by parts, some of the total derivative terms are removed from action (\ref{A2}) and one is left with,
\be\begin{split}\label{15}
A_2 &= \int\Big[\frac {1}{16 \pi G}\left( - \frac{3 {\dot z}^2}{2 N \sqrt z} + 6 k N \sqrt z \right) + \frac{9 \xi \beta}{\sqrt z}\left(\frac{{\ddot z}^2}{N^3} - \frac{2 \dot N \dot z \ddot z}{N^4} + \frac{{\dot N}^2{\dot z}^2}{N^5} + \frac{2 k {\dot z}^2}{N z} + 4 k^2 N \right) \\
&-\frac{36\beta k\xi'\dot z\dot\phi}{N\sqrt z} - \frac {\xi^{'} \dot z \dot {\phi}}{N \sqrt z}\left(\frac{{\dot z}^2}{N^2 z} + 12 k\right) + z^{\frac{3}{2}}\left(\frac {1}{2N} {\dot {\phi}}^2 - VN\right)\Big]dt +\xi(\phi)\beta \Sigma_{R^2_2},
\end{split}\ee
At this stage, following Horowitz \cite{horo}, let us introduce an auxiliary variable
\be\label{Q1} Q = \frac{\partial A}{\partial \ddot z} = \frac{18\xi\beta}{N^3 \sqrt z}\left(\ddot z  - \frac{\dot N \dot z}{N}\right).\ee
judiciously in action (\ref{15}) as,
\be\begin{split} \label{17}
A_2 &= \int\Big[\frac {1}{16 \pi G}\left( - \frac{3 {\dot z}^2}{2 N \sqrt z} + 6 k N \sqrt z \right) + Q \ddot z - \frac{\dot N}{N}\dot z Q - \frac{N^3 \sqrt z}{36\xi \beta} Q^2 + \frac{18\xi \beta k}{\sqrt z}\left(\frac{{\dot z}^2}{N z} + 2 k N \right) \\
&-\frac{36\beta k\xi'\dot z\dot\phi}{N\sqrt z} - \frac {\xi^{'} \dot z \dot {\phi}}{N \sqrt z}\left(\frac{{\dot z}^2}{N^2 z} + 12 k\right) + z^{\frac{3}{2}}\left(\frac {1}{2N} {\dot {\phi}}^2 - VN\right)\Big]dt + \xi(\phi)\beta\Sigma_{R^2_2},
\end{split}.\ee
Under integration by parts, the rest of the boundary terms gets cancelled with the total derivative term, and the action (\ref{17}) is finally expressed as,
\be\begin{split}\label{55}
A_2 &= \int\Big[\frac {1}{16 \pi G}\left( - \frac{3 {\dot z}^2}{2 N \sqrt z} + 6 k N \sqrt z \right) - \dot Q \dot z - \frac{\dot N}{N}\dot z Q - \frac{N^3 \sqrt z}{36 \beta\xi} Q^2 + \frac{18 \beta\xi k}{\sqrt z}\left(\frac{{\dot z}^2}{N z} + 2 k N \right) \\
&-\frac{36\beta k\xi'\dot z\dot\phi}{N\sqrt z} - \frac {\xi^{'}\dot z \dot {\phi}}{N \sqrt z}\left(\frac{{\dot z}^2}{N^2 z} + 12 k\right) + z^{\frac{3}{2}}\left(\frac {1}{2N} {\dot {\phi}}^2 - VN\right)\Big]dt.
\end{split}\ee
The canonical momenta are,
\begin{subequations}\begin{align}
& p_Q = - \dot z \label{pQ} \\
& p_z = -\frac{3\dot z}{16 \pi G N \sqrt z} - \dot Q - \frac{Q \dot N}{N} +\frac{36 \beta k\xi \dot z}{N z^{\frac {3}{2}}}-\frac{36\beta k\xi'\dot\phi}{N\sqrt z}- \frac{3\xi' \dot\phi}{N \sqrt z}\left(\frac{\dot z^2}{N^2 z} + 4 k\right) \label{12}\\
& p_\phi = -\frac{36\beta k\xi'\dot z}{N\sqrt z}- \frac{\xi' \dot z}{N \sqrt z}\left(\frac{\dot z^2}{N^2 z} + 12 k\right) + \frac{z^{\frac {3}{2}}}{N} \dot\phi \label{13} \\
& p_N = -\frac{Q \dot z}{N}
\end{align}\end{subequations}
The $N$ variation equation reads
\be\begin{split} \label{23}
& - \frac {3}{16 \pi G}\left(\frac{{\dot z}^2}{2 N^2 \sqrt z} + 2 k \sqrt z \right) - \frac{Q \ddot z}{N}- \frac{\dot Q \dot z}{N} + \frac{N^2 \sqrt z}{12 \beta\xi} Q^2 + \frac{18 \xi\beta k}{\sqrt z}\left(\frac{{\dot z}^2}{N^2 z} - 2 k\right)-\frac{36\beta k\xi'\dot z\dot\phi}{N^2\sqrt z} - \frac {3 \xi^{'} \dot z \dot {\phi}}{N^2 \sqrt z}\left(\frac{{\dot z}^2}{N^2 z} + 4 k\right) \\
& + z^{\frac{3}{2}}\left(\frac {1}{2N^2} {\dot {\phi}}^2 + V\right) = 0.
\end{split}\ee
Action (\ref{55}) is singular due to the diffeomorphic invariance. This must be reflected in the $N$ variation equation, since it should not contain second derivative term. Presence of the second derivative term ($\ddot z$) indicates that the situation is altogether different from GTR. This awful situation may be handled easily without going through Dirac's constraint analysis. This is possible because, in view of the definition of $Q$ given in (\ref{Q1}), $\ddot z$ term may be removed from the above $N$ variation equation (\ref{23}), which now takes the form,
\be\begin{split}\label{a}
&- \frac {3}{16 \pi G}\left(\frac{{\dot z}^2}{2 N^2 \sqrt z} + 2 k \sqrt z \right) - \frac{\dot z \dot Q}{N} - \frac{\dot N\dot z Q}{N^2} + \frac{N^2 \sqrt z}{36 \beta\xi}Q^2  + \frac{18 \xi\beta k}{\sqrt z}\left(\frac{{\dot z}^2}{N^2 z} - 2 k\right)-\frac{36\beta k\xi'\dot z\dot\phi}{N^2\sqrt z} - \frac {3 \xi^{'} \dot z \dot {\phi}}{N^2 \sqrt z}\left(\frac{{\dot z}^2}{N^2 z} + 4 k\right) \\
& + z^{\frac{3}{2}}\left(\frac {1}{2N^2} {\dot {\phi}}^2 + V\right) = 0.
\end{split}\ee
Now, since the above equation \eqref{a} does not contain second derivative term, so it should be treated is a constraint of the system. It can be easily verified that this is the Hamiltonian of the system in disguise. We can therefore write Hamiltonian constraint equation as,
\be\begin{split}\label{1}
H_c &= N \Bigg[ - \frac {3}{16 \pi G}\left(\frac{{\dot z}^2}{2 N^2 \sqrt z} + 2 k \sqrt z \right) - \frac{\dot z \dot Q}{N} - \frac{\dot N\dot z Q}{N^2} + \frac{N^2 \sqrt z}{36 \beta\xi}Q^2  + \frac{18 \xi\beta k}{\sqrt z}\left(\frac{{\dot z}^2}{N^2 z} - 2 k\right)-\frac{36\beta k\xi'\dot z\dot\phi}{N^2\sqrt z}\\& - \frac {3 \xi^{'} \dot z \dot {\phi}}{N^2 \sqrt z}\left(\frac{{\dot z}^2}{N^2 z} + 4 k\right) + z^{\frac{3}{2}}\left(\frac {1}{2N^2} {\dot {\phi}}^2 + V\right)\Bigg] = 0
\end{split}\ee
which is constrained to vanish. Now, using the expression,
\be p_Q p_z = \frac{3\dot z^2}{16\pi G N \sqrt z} + \dot z\dot Q + \frac{\dot N}{N}\dot z Q - \frac{36 k \beta\xi \dot z^2}{N z^{\frac{3}{2}}} +\frac{36\beta k\xi'\dot z\dot\phi}{N\sqrt z}+ \frac{3 \dot z \xi^{'} \dot\phi}{N \sqrt z}\left(\frac{\dot z^2}{N^2 z} + 4k\right)\ee
and also using the definitions of momenta (\ref{pQ}, \ref{13}), the Hamiltonian constraint equation in terms of the phase space variables is expressed as,

\be\begin{split}\label{2}
H_c &= \frac{3}{16\pi G}\Big(\frac{{p_Q}^2}{2 N \sqrt z} - 2k N \sqrt z \Big)-p_Q p_z+ \frac{N^3 Q^2 \sqrt z}{36\beta\xi} - \frac{18 \xi k \beta}{\sqrt z} \Big(\frac{{p_Q}^2}{N z} + 2k N\Big) +\frac{N{p_\phi}^2}{2z^{\frac{3}{2}}} + \frac {{\xi^{'}}^2 {p_Q}^6}{2 N^5 z^{\frac{9}{2}}} +\frac{648k^2\beta^2\xi'^2p_Q^2}{Nz^{\frac{5}{2}}} \\
& + \frac {72k^2 {\xi^{'}}^2 {p_Q}^2}{N z^{\frac{5}{2}}}-\frac{36\beta k\xi' p_Qp_{\phi}}{z^2}+ \frac {12k {\xi^{'}}^2 {p_Q}^4}{N^3 z^{\frac{7}{2}}} - \frac {\xi^{'} {p_Q}^3 p_\phi}{N^2 z^3} - \frac {12 k \xi^{'}p_Q p_\phi}{z^2}+\frac{36\beta k\xi'^2p_Q^4}{N^3z^{\frac{7}{2}}}+\frac{432\beta k^2\xi'^2p_Q^2}{Nz^{\frac{5}{2}}}\\& + N V z^{\frac{3}{2}}= 0 .
\end{split}\ee
In the above form of phase-space Hamiltonian \eqref{2}, various momenta appear in products whose powers are at least of second order and reach up to sixth order (${p_Q}^6$). Of course, this is very inconvenient in order to form the operators. Even if one does, a large number of initial (boundary) conditions are required to solve the quantum counterpart, which are not available. Further, Hamiltonian \eqref{2} also contains cross terms in higher degree of the momenta (${p_Q}^3 p_{\phi}$), which makes thing even complicated. On the contrary, one can observe that the configuration variable $Q$ appears only quadratically in \eqref{2}. So, in order to handle such awful situation, it is suggestive to express the above Hamiltonian in terms of the basic variables $\{K_{ij}, ~\pi^{ij}\}$. This is possible under canonical transformation from $\{Q, p_Q\}$ to $\{x, p_x\}$, as $Q = \frac{p_x}{N}$ and $ p_{Q} = -\dot z = -Nx $. Hence, the Hamiltonian in terms of the basic variables reads,

\be\begin{split}\label{3}
H_c &= N\Big[x p_z + \frac{{p_x}^2 \sqrt z}{36\beta\xi} + \frac{{p_\phi}^2}{2z^{\frac{3}{2}}} +\frac{\xi'}{z^2}\left(36\beta kx+ \frac{ x^3}{z} + 12 kx \right)p_\phi  + \frac{3}{16\pi G}\Big(\frac{x^2}{2 \sqrt z} - 2k \sqrt z \Big) - \frac{18 \xi k \beta}{\sqrt z} \Big(\frac{x^2}{z} + 2k\Big) \\&+ \frac {{\xi^{'}}^2 x^6}{2 z^{\frac{9}{2}}}+\frac{648k^2\beta^2\xi'^2x^2}{z^{\frac{5}{2}}}  + \frac {72k^2{\xi^{'}}^2 x^2}{z^{\frac{5}{2}}} + \frac {12k {\xi^{'}}^2 x^4}{z^{\frac{7}{2}}}+\frac{36\beta k\xi'^2x^4}{z^{\frac{7}{2}}}+\frac{432\beta k^2\xi'^2x^2}{z^{\frac{5}{2}}} + V z^{\frac{3}{2}}\Big]= N{\mathcal{H}} = 0.
\end{split}\ee
The fact that the above form (\ref{3}) gives the correct Hamiltonian description of the theory (\ref{A2}) under consideration, has been established in the appendix. The above description of the phase-space Hamiltonian explores the very importance of basic variables $\{h_{ij},p_{ij}; K_{ij}, \pi^{ij}\}$, which are $\{z, p_z; x,p_x\}$ respectively in the minisuperspace under consideration. First of all, the diffeomorphic invariance $H_c = N{\mathcal H}$, being an artefact of general covariance is manifest in GTR. In higher-order theory of gravity, as one can observe, it is manifest only when the Hamiltonian is expressed in terms of the basic phase-space variables. Arbitrary phase-space variables, even if those are related to the basic variables under canonical transformation, are no good. Next, the above Hamilton constraint equation \eqref{3} is only quadratic in momenta, and there are no cross terms. Thus, it is now convenient to canonically quantize the Hamilton constraint equation \eqref{3}. Moreover, the momentum $p_{z}$ appears only linearly, which is nice to realize the internal variable $z$ as time co-ordinate, upon quantization.\\

One may wonder that, why basic variables were not used from the vary beginning? The answer to the question is, if the action were expressed in terms of $\{h_{ij}, K_{ij}\}$ from the very beginning, the Hessian determinant would have vanished and the Lagrangian would have been singular. Such situation may be dealt with Dirac's constraint analysis \cite{dirac}. Nevertheless, Dirac's analysis, implicitly assumes that $h_{ij}$ and $K_{ij}$ should be kept fixed at the boundary. As a result, supplementary boundary terms are not required. In the process, one looses the cherished Gibbons-Hawking-York (GHY) boundary term \cite{ghy}, which is associated with the entropy of a black hole. Even in the weak field approximation ($\xi \rightarrow 0$), it can't be retrieved. On the contrary, it has been shown that the supplementary boundary term associated with higher order theory of gravity reproduces the expected ADM energy \cite{ADM} upon passing to the Hamiltonian formalism, and the correct expression of entropy of a Schwarzschild black hole may be found in the semiclassical limit \cite{DH}. In the present analysis therefore, along with $h_{ij}$, the Ricci scalar $R$ has been kept fixed at the boundary, and the action is supplemented by appropriate boundary terms. Auxiliary variable has been introduced so that the Lagrangian is non-singular. In this context, we also mention that, the two techniques yield different phase-space Hamiltonian in general, as has been demonstrated recently \cite{aks1a}.\\

It is now straightforward to write the action in the ADM canonical form as,
\be \label{CA}A_2 = \int (\dot z p_z + \dot x p_x + \dot\phi p_\phi - N\mathcal{H})~dt~ d^3x = \int (\dot h_{ij} p^{ij} + \dot K_{ij} \pi^{ij} + \dot\phi p_\phi - N\mathcal{H})~dt~ d^3x.\ee
Thus, the problem of branching, which appeared due to the presence of higher degree term in the action (\ref{A1}) has been bypassed by the introduction of a higher order curvature invariant term - $R^2$. Note that such a term appears in the weak energy limit of all different quantum theory of gravity, and so the action (\ref{A2}) is a natural generalization of action (\ref{A1}). In fact, any higher order curvature invariant term can cure the problem associated with higher degree, but as $R_{\mu\nu} R^{\mu\nu}$ leads to ghost, so it is safe to handle the situation, with scalar curvature invariant term.

\subsection{Canonical Quantisation}
The quantum version of Hamiltonian (\ref{3}) is
\be\begin{split}\label{71}
\frac{i\hbar}{\sqrt z}\frac{\partial \Psi}{\partial z} &= -\frac{\hbar^2}{36\beta\xi x}\left(\frac{\partial^2}{\partial x^2} + \frac{n}{x}\frac{\partial}{\partial x}\right)\Psi - \frac{\hbar^2}{2x z^2}\frac{\partial^2\Psi}{\partial \phi^2} + \frac{1}{z^{\frac{5}{2}}}\left(\frac {x^2}{z}+36\beta k + 12 k\right)\widehat{\xi'}\widehat{p}_\phi + \frac{3}{16\pi G}\left(\frac{x}{2 z} - \frac{2k}{x}\right)\Psi\\& - \frac{18 k \beta\xi}{z} \left(\frac{x}{z} + \frac{2k}{x}\right)\Psi+\frac{\widehat{\xi'}^2 x}{z^3}\left(\frac {x^4}{2 z^2} + \frac {12k x^2}{z} + 72k^2+648k^2\beta^2+\frac{36\beta k x^2}{z}+432 k^2 \beta\right)\Psi +\frac{ V z}{x}\Psi \\&= \hat H_e\Psi,
\end{split}\ee
where, $n$ is the operator ordering index. Operator form of $\widehat\xi' \widehat p_{\phi}$ appearing on the third term on the right hand side, may be inserted appropriately, only after knowing the specific form of $\xi(\phi)$, as this term also requires ordering. Specific form of $\xi(\phi)$ is also required to investigate the behaviour of the quantum theory, under certain appropriate semi-classical approximation. A specific form of $\xi(\phi)$ and $V(\phi)$ may be obtained if we invoke slow roll inflation, so that almost scale-invariant perturbation on large scales is successfully generated from quantum fluctuations of $\phi$. In view of the action \eqref{55}, the $(^0_0)$ equation of Einstein (which is the essentially the Hamiltonian constraint equations (\ref{1}), (\ref{2}) or (\ref{3})) and the $\phi$ variation equation may be expressed under the standard gauge choice $N = 1$, and setting $k = 0$, in terms of the scale factor $a$ as

\be\begin{split}\label{phiv}\frac{\dot a^2}{a^2} &=  - 96\beta \xi\pi G\left[2\frac{\dot a\dddot a}{a^2} - \frac{\ddot a^2}{a^2} + 2\frac{\dot a^2\ddot a}{a^3} - 3\frac{\dot a^4}{a^4} \right] - 192\pi G\beta\xi'\dot\phi\left(\frac{\dot a\ddot a}{a^2}+\frac{\dot a^3}{a^3}\right)-64\pi G\xi'\dot\phi \left(\frac{\dot a^3}{a^3}\right) + \frac{8\pi G}{3}\left(\frac{\dot\phi^2}{2} + V\right)\end{split},\ee
and
\be\label{clfld5}
 - 24\xi'\dot a^2\ddot a-36\beta\xi'a\ddot a^2-72\beta\xi'\dot a^2\ddot a-36\beta\xi' \frac{\dot a^4}{a} + 3a^2\dot a\dot\phi + a^3(\ddot\phi + V') = 0,\ee
respectively. The above equations (\ref{phiv}), (\ref{clfld5}) may further be rearranged as
\be\begin{split}\label{phi1} M_{pl}^2 H^2 &=\frac{1}{3}\left(\frac{1}{2}\dot\phi^2+V\right)-{12\beta\xi}\left[4H^4\left(1+\frac{\dot H}{H^2}\right)+4H^2\dot H\left(1+\frac{\dot H}{H^2}\right)+2H^3\left(\frac{\ddot H}{H^2}-2\frac{\dot H^2}{H^3}\right) -H^4\left(1+\frac{\dot H}{H^2}\right)^2-3H^4\right]\\&-{48\beta}\dot\xi H^3-{8}\dot\xi H^3-{24}H\dot\xi \dot H,\end{split}\ee
\be\label{hc}
\ddot\phi + 3H\dot\phi=-V'+\left(6\beta+1\right)24\xi'H^4+\left(6\beta+1\right)24\xi'H^2\dot H+36\beta\xi'\dot H^2,\ee
where $H\equiv\frac{\dot a}{a}$ denotes the expansion rate and $M_{pl}^2=\frac{1}{8\pi G}$.
Since the GB coupling is a function of $\phi,$ one has $\dot\xi=\xi'\dot\phi$ and $\ddot\xi=\xi''\dot\phi^2+\xi'\ddot\phi$.
Now, due to the presence of an additional degree of freedom $\xi(\phi)$, along with the standard slow-roll conditions of minimally coupled single-field inflation, viz. $\dot\phi^2\ll V$ and $|\ddot\phi|\ll 3H|\dot\phi|$, it is required to impose two additional conditions, viz. $4|\dot \xi| H \ll 1$ and $|\ddot\xi| \ll |\dot\xi| H$ \cite{addition}. Instead of standard slow roll parameters, it is customary to introduce a combined hierarchy of Hubble and Gauss-Bonnet flow parameters \cite{15}. Firstly, the background evolution is described by a set of horizon flow functions (the behaviour of Hubble distance during inflation) starting from

\be \label{hf1}\epsilon_0 = {d_H\over d_{H_i}}, ~~\mathrm{where,}~d_H=H^{-1}\ee
is the Hubble distance, also called horizon in our chosen unit. Now hierarchy of functions is defined in a systematic way as
\be\label{hf2}\epsilon_{l+1} = {d\ln |\epsilon_l|\over d N}, ~~l\ge 0,\ee
In view of the definition $N = \ln {a\over a_i}$, which implies $\dot N = H$, one can compute $\epsilon_1 = {d\ln d_H\over d N}$ , which is the logarithmic change of Hubble distance per e-fold expansion $N$, and is the first slow-roll parameter $\epsilon_1 = \dot d_H = -{\dot H\over H^2}$. The above hierarchy allows one to compute $\epsilon_2 = {d\ln \epsilon_1\over d N} = {1\over H}{\dot\epsilon_1\over \epsilon_1}$, which implies $\epsilon_1\epsilon_2 = d_H\ddot d_H = -{1\over H^2}\Big({\ddot H\over H}-2{\dot H^2\over H^2}\Big)$. In the same manner higher slow-roll parameters may be computed. Equation \eqref{hf2} essentially defines a flow in space with cosmic time being the evolution parameter, which is described by the equation of motion
\be\label{hf3} \epsilon_0\dot\epsilon_l -{1\over d_{H_i}} \epsilon_l\epsilon_{l-1} = 0, ~~l\ge 0.\ee
One can also check that \eqref{hf3} yields all the results obtained from the hierarchy defined in \eqref{hf2}, using definition \eqref{hf1}. As already mentioned, the additional degree of freedom appearing due to the Gauss-Bonnet-Dilatonic coupling, requires to introduce yet another hierarchy of Gauss-Bonnet flow parameters as

\be \label{hf4} \delta_1 = 4\dot \xi H \ll 1,\;\;\;\delta_{i+1} =  {d\ln |\delta_i|\over d\ln a}, ~\mathrm{with},~ i \ge 1.\ee
Clearly for $i = 1$, $\delta_2 = {d\ln|\delta_1|\over d N} = {1\over \delta_1}{\dot\delta_1\over \dot N}$, and $\delta_1\delta_2 = {4\over H}(\ddot \xi H + \dot\xi\dot H)$ and so on. The slow-roll conditions therefore read $|\epsilon_i|\ll 1$ and $|\delta_i|\ll1$, which is analogous to the standard slow-roll approximation, and the above equations \eqref{phi1} and \eqref{hc} may therefore be expressed as
\be\begin{split} \label{f1} M_{pl}^2 H^2 =&\frac{1}{3}\left(\frac{1}{2}\dot\phi^2+V\right)-{12\beta\xi}\left[3H^4\left(1-\epsilon_1\right)^2 +2H^3\dot{\overbrace{(1-\epsilon_1)}}-3H^4 \right]-\left(2H^2+12\beta H^2\right)\left(1+\delta_1\right)\\&+6\beta H^2\left(1+\delta_1\epsilon_1\right)+\left(2H^2+6\beta H^2\right)\end{split} \ee
and
\be \begin{split}\label{f2} \ddot\phi+3H\dot\phi=&-V'+\frac{1}{\dot\phi} \left[(6H^3\left(6\beta+1\right)\left(1+\delta_1\right)-6H^3\left(6\beta+1\right)\left(1+\delta_1\epsilon_1\right)
+9H^3\beta\left(1+\delta_1\right)\left(1-\epsilon_1\right)^2\right] \\&
+\frac{9\beta H^3}{\dot\phi}\left(\left(1+\delta_1\right)-\left(1-\epsilon_1\right)^2-1\right) \end{split}\ee
respectively. In view of the slow-roll parameters, the above equations \eqref{f1}, and \eqref{f2} may be approximated to
\be\label{H2} H^2\simeq \frac{1}{3M_{pl}^2}V,\ee
\be\label{HH}H\dot\phi\simeq -\frac{1}{3}V \mathcal{Q}, \ee
where, $\mathcal{Q}=\frac{V'}{V}$. In deriving equation \eqref{HH}, the approximation arrived at equation \eqref{H2} has been used. The number of e-folds may then be computed as usual in view of the following relation,
\be\label{ef}N(\phi)\simeq\int^{t_f}_t H dt=\int^{\phi_f}_{\phi}\frac{H}{\dot\phi}d\phi\simeq\frac{1}{M_{pl}^2}\int^{\phi}_{\phi_f}\frac{d\phi}{\mathcal{Q}}\ee
where, $\phi$ and $\phi_f$ denote the values of the scalar field at the beginning ($t$) and the end ($t_f$) of inflation. Let us now consider a specific model with a monomial potential and an inverse monomial GB coupling as,
\be V(\phi)=V_1+V_0\phi^m ,~~~~~~ \xi(\phi)=\xi_0\phi^{-m}\ee
where, $V_0$, $V_1$, $\xi_0$ and $m$ are constants. Under the choice $m=2$, $\mathcal{Q}$ may be expressed as,
\be \mathcal{Q}= {V'\over V} = \frac{2V_0\phi}{V_1+V_0\phi^2}. \ee
Therefore the number of e-folding \eqref{ef} reads
\be N(\phi)=\frac{1}{2M_{pl}^2}\left[{\frac{V_1}{V_0}}{\ln{\left(\frac{\phi}{\phi_f}\right)}}+\frac{
\left(\phi^2-\phi_f^2\right)}{2}\right]\ee
Now if we set $\frac{V_1}{V_0}\approx {-6M_{pl}^2},$ then $N(\phi)\gtrsim 60,$ for $\phi\gtrsim 16.4 M_{pl}$ and $\phi_f\sim 3.26 M_{pl}$ which solves the horizon and flatness problem. Slow roll ends under the condition $ \epsilon_1 = \frac{M_{pl}^2}{2}{\left(V'\over V\right)^2} = \frac{2 M_{pl}^2\phi^2}{\left({V_1\over V_0} +\phi^2\right)^2}\geqslant 1$, which requires $\phi_f \lesssim 3.26 M_{pl}$. Cosmological perturbation with dilatonic Gauss-Bonnet gravity has been studied extensively in the literature \cite{gb1}. The presence of an additional $R^2$ term doesn't seem to alter the qualitative behaviour. At present let us therefore concentrate on the results in the quantum domain, which is our main concern. Under the above choice of $m = 2$ and $k=0$, the quantum equation (\ref{71}) may now be expressed as
\be\begin{split}\label{7}
\frac{i\hbar}{\sqrt z}\frac{\partial \Psi}{\partial z} &= -\frac{\hbar^2}{36\beta\xi x}\left(\frac{\partial^2}{\partial x^2} + \frac{n}{x}\frac{\partial}{\partial x}\right)\Psi - \frac{\hbar^2}{2x z^2}\frac{\partial^2\Psi}{\partial \phi^2} - i\hbar \frac{\xi_0 x^2}{z^{\frac{7}{2}}}\left(\frac{3 \Psi}{\phi^4} - \frac{2}{\phi^3}\frac{\partial\Psi }{\partial \phi}\right) + \left[\frac{3x}{32\pi Gz} + \frac{2\xi_0^2 x^5}{\phi^6z^5}+\frac{ V z}{x} \right]\Psi,
\end{split}\ee
where Weyl symmetric ordering has been performed in the third term appearing on right hand side. Now, again under a further change of variable, the above modified Wheeler-de-Witt equation, takes the look of Schr\"odinger equation, viz.,

\be\begin{split}\label{5}
i\hbar\frac{\partial \Psi}{\partial \alpha} &= -\frac{\hbar^2}{54\beta\xi}\left(\frac{1}{x}\frac{\partial^2}{\partial x^2} + \frac{n}{x^2}\frac{\partial}{\partial x}\right)\Psi - \frac{\hbar^2}{3x \alpha^{\frac{4}{3}}}\frac{\partial^2\Psi}{\partial \phi^2} - i\hbar \frac{2\xi_0 x^2}{3\alpha^{\frac{7}{3}}}\left(\frac{3\Psi}{\phi^4} - \frac{2}{\phi^3}\frac{\partial \Psi}{\partial\phi}\right) + V_e\Psi = \hat H_e\Psi,
\end{split}\ee
where, $\alpha  = z^{\frac{3}{2}} = a^3$ plays the role of internal time parameter. In the above, the effective potential $V_e$, is given by,
\be\label{Ve} V_e = \frac{x}{16\pi G\alpha^{\frac{2}{3}}} + \frac{4\xi_0^2 x^5}{3\phi^6\alpha^{\frac{10}{3}}}+ \frac{2V \alpha^{\frac{2}{3}}}{3x}. \ee
The hermiticity of $\hat H_e$ should enable to write the continuity equation, which requires to find ${\partial \rho\over\partial\alpha}$, where, $\rho = \Psi^*\Psi$. Little algebra leads to the following equation,
\be \begin{split}
{\partial \rho\over\partial\alpha} &= -{\partial \over\partial x}\left[\frac{i \hbar }{54\beta \xi x}(\Psi\Psi^*_{,x}-\Psi^*\Psi_{,x})\right] - {\partial \over\partial \phi}\left[\frac{i \hbar }{3\alpha^{\frac{4}{3}}x}(\Psi\Psi^*_{,\phi}-\Psi^*\Psi_{,\phi})-\frac{4\xi_0 x^2 }{3\alpha^{\frac{7}{3}}\phi^3}\Psi^*\Psi\right]\\
& + \frac{(n+1)}{x^2}\left(\Psi\Psi^*_{,x}-\Psi^*\Psi_{,x}\right).\end{split}\ee
Clearly, continuity equation can be written, only under the choice $n=-1$, as

\be \frac{\partial\rho}{\partial \alpha} + \nabla . {\bf{J}} = 0. \ee
In the above, $\rho = \Psi^*\Psi$ and ${\bf J} = ({\bf J}_x, {\bf J}_\phi, 0)$ are the probability density and the current density respectively, where
\begin{subequations}\begin{align}
{\bf J}_x &= \frac{i \hbar }{54\beta \xi x}(\Psi\Psi^*_{,x}-\Psi^*\Psi_{,x})\\
{\bf J}_\phi &= \frac{i \hbar }{3x\alpha^{\frac{4}{3}}}(\Psi\Psi^*_{,\phi}-\Psi^*\Psi_{,\phi}) - \frac{4\xi_0 x^2 }{3\alpha^{\frac{7}{3}}\phi^3}\Psi^*\Psi
\end{align}
\end{subequations}
In the process, operator ordering index has been fixed as $n = -1$ from physical argument. As already mentioned, here the variable $\alpha$ plays the role of internal time parameter. It is important to note that, standard quantum mechanical probabilistic interpretation of the theory has been possible taking $\alpha = z^{3\over 2} = a^3$, i.e. the proper volume as the internal time parameter.

\subsection{Classical and Semiclassical solutions (under WKB approximation)}

To check the viability of the quantum equation (\ref{5}), so obtained, it is required to test its behaviour under certain appropriate semi-classical approximation. Semiclassical approximation should be performed with the full quantum equation, without assuming slow roll conditions. This requires a viable classical solution of the field equations \eqref{phi1} and \eqref{hc}. Under the choice $V =V_1+ V_0 \phi^2, \xi =\xi_0\phi^{-2}$, where $V_1$ is a constant, the full classical field equations admit exponential inflationary solution in the form
\be\label{11} ~~~~~~a = a_0 e^{\mathrm {H} t}~~~~~~\phi=\phi_0 e^{-\mathrm {H} t}\ee
restricting the constants to, $V_1 = \frac{3\mathrm {H}^2}{8\pi G}, \beta = - {1\over 6}~\text{and}~V_0 = -{\mathrm {H}^2\over 2},$ where $\mathrm {H}$ is yet another constant. Now, let us express equation (\ref{7}) as,
\be\begin{split}\label{8}
-\frac{\hbar^2\sqrt z}{36\beta\xi x}\left(\frac{\partial^2}{\partial x^2} + \frac{n}{x}\frac{\partial}{\partial x}\right)\Psi - \frac{\hbar^2}{2x z^{\frac{3}{2}}}\frac{\partial^2\Psi}{\partial \phi^2} - i\hbar\frac{\partial \Psi}{\partial z} + i\hbar \frac{2\xi_0x^2}{z^3\phi^3}\frac{\partial \Psi}{\partial \phi} + \mathcal{V}\Psi = 0
\end{split}\ee
where
\be \mathcal{V} = \frac{3x}{32\pi G\sqrt{z}} + \frac{2{\xi_0}^2 x^5}{\phi^6z^{\frac{9}{2}}} - \frac{3i\hbar \xi_0x^2}{z^3\phi^4} + \frac{V_0\phi^2z^{\frac{3}{2}}}{x} + \frac{V_1z^{\frac{3}{2}}}{x}.\ee
The above equation may be treated as time independent Schr{\"o}dinger equation with three variables $x$, $z$ and $\phi$ therefore, as usual, let us sought the solution of equation (\ref{8}) as,
\be\label{9} \psi = \psi_0e^{\frac{i}{\hbar}S(x,z,\phi)}\ee
and expand $S$ in power series of $\hbar$ as,
\be\label{10} S = S_0(x,z,\phi) + \hbar S_1(x,z,\phi) + \hbar^2S_2(x,z,\phi) + .... \ .\ee
Now inserting the expressions (\ref{9}) and (\ref{10}) in equation (\ref{8}) and equating the coefficients of different powers of $\hbar$ to zero, one obtains the following set of equations (upto second order)
\begin{subequations}\begin{align}
&\frac{\sqrt z}{36\beta \xi x}S_{0,x}^2 + \frac{S_{0,\phi}^2}{2xz^{\frac{3}{2}}} + S_{0,z} - \frac{2\xi_0 x^2}{z^3\phi^3} S_{0,\phi} +  \frac{3x}{32\pi G\sqrt{z}} + \frac{2{\xi_0}^2 x^5}{\phi^6z^{\frac{9}{2}}} + \frac{V_0\phi^2z^{\frac{3}{2}}}{x} + \frac{V_1z^{\frac{3}{2}}}{x}  = 0 \label{hbar0} \\
& -\frac{i\sqrt z}{36\beta \xi x}S_{0,xx} - \frac{in\sqrt z}{36\beta \xi x^2}S_{0,x} - \frac{iS_{0,\phi\phi}}{2xz^{\frac{3}{2}}} + S_{1,z} + \frac{\sqrt zS_{0,x}S_{1,x}}{18\beta \xi x} + \frac{2S_{0,\phi}S_{1,\phi}}{xz^{\frac{3}{2}}} + \frac{\xi_0x^2}{z^3\phi^3} S_{1,\phi} - \frac{3i \xi_0x^2}{z^3\phi^4} = 0. \label{hbar1}\\
&  - i\frac{\sqrt z S_{1,xx}}{36\beta \xi x} + \frac{\sqrt z {S_{1,x}}^2}{36\beta \xi x} + \frac{\sqrt z S_{0,x}S_{2,x}}{18\beta\xi x} - i\frac{n\sqrt zS_{1,x}}{36\beta\xi x^2} - i\frac{S_{1,\phi\phi}}{2xz^{\frac{3}{2}}} + \frac{{S_{1,\phi}}^2}{2xz^{\frac{3}{2}}}+ \frac{2S_{0,\phi}S_{2,\phi}}{xz^{\frac{3}{2}}} + S_{2,z} - \frac{\xi_0x^2}{z^3\phi^3} S_{2,\phi} = 0,
\end{align}\end{subequations}
which are to be solved successively to find $S_0(x,z,\phi),\; S_1(x,z,\phi)$ and $S_2(x,z,\phi)$ and so on. Now identifying $S_{0,x}$ as $p_x$; $S_{0,z}$ as $p_z$ and $S_{0,\phi}$ as $p_{\phi}$ one can recover the classical Hamiltonian constraint equation $H_c = 0$ (for $k=0$), given in equation (\ref{3}) from equation (\ref{hbar0}). Thus, $S_{0}(x, z)$ can now be expressed as,
\be\label{14} S_0 = \int p_z dz + \int p_x dx + \int p_\phi d\phi \ee
apart from a constant of integration which may be absorbed in $\psi_0$. Clearly, $S_0$ may only be found, evaluating the above integral, which is impossible due to tight coupling. We therefore use the classical solution (\ref{11}) for the purpose, and relate the (once) independent variables $a$ and $\phi$ through their temporal function as $z=a^2=\frac{(a_0\phi_0)^2}{\phi^2}$, which is always possible (remember that to find the trajectory of a parabola, we relate the co-ordinates $x$ and $y$, through their time dependence). The integrals in the above expression can now be evaluated using the definition of $p_z$ given in (\ref{12}), $p_{\phi}$ in (\ref{13}) and $p_x = N Q$, and also recalling the expression for $Q$ given in (\ref{Q1}). Further, we choose $n = -1$, since probability interpretation holds only for such value of $n$. Hence, $x (= \dot z), ~\phi$ and the expressions of $p_x$, $p_z$ $p_\phi$ can be expressed in terms of $z$ and $\phi$ as,
\begin{subequations}\begin{align}
&\label{comb} x = 2{\mathrm H} z,~~ \phi = {a_0\phi_0\over \sqrt z} \\
&p_x = - \frac{12\xi_0 {\mathrm H}^2z^{\frac{3}{2}}}{(a_0\phi_0)^2} = - \frac{3\xi_0 \sqrt{2\mathrm H}x^{\frac{3}{2}}}{(a_0\phi_0)^2}\\
&p_z = -\frac{3}{8\pi G}\sqrt z{\mathrm H} + \frac{12\xi_0 {\mathrm H}^3 z^{\frac{3}{2}}}{(a_0\phi_0)^2} \\
&p_\phi = 16\xi_0 {\mathrm H}^3\frac{(a_0\phi_0)^3}{\phi^6} - \mathrm H\frac{(a_0\phi_0)^3}{\phi^2}
\end{align}\end{subequations}
and the integrals in (\ref{14}) are evaluated as,
\begin{subequations}\begin{align}
&\int p_x dx = - \frac{6\xi_0 \sqrt{2\mathrm H}x^{\frac{5}{2}}}{5(a_0\phi_0)^2} = - \frac{48\xi_0 {\mathrm H}^3z^{\frac{5}{2}}}{5(a_0\phi_0)^2}; \\
&\int p_z dz =  - \frac{1}{4\pi G}\mathrm{H} z^{\frac{3}{2}} + \frac{24\xi_0 {\mathrm H}^3 z^{\frac{5}{2}}}{5(a_0\phi_0)^2}; \\
&\int p_\phi d\phi = - \frac{16\xi_0 {\mathrm H}^3(a_0\phi_0)^3}{5\phi^5} + \mathrm H\frac{(a_0\phi_0)^3}{\phi}.
\end{align}\end{subequations}
\noindent
At this end, explicit form of $S_0$ can be written as,
\be\begin{split}\label{S0} S_0 &= - \frac{6\xi_0 \sqrt{2\mathrm H}x^{\frac{5}{2}}}{5(a_0\phi_0)^2}- \frac{1}{4\pi G}\mathrm{H} z^{\frac{3}{2}} + \frac{24\xi_0 {\mathrm H}^3 z^{\frac{5}{2}}}{5(a_0\phi_0)^2}- \frac{16\xi_0 {\mathrm H}^3(a_0\phi_0)^3}{5\phi^5} + \mathrm H\frac{(a_0\phi_0)^3}{\phi}\\
&= - \frac{1}{4\pi G}\mathrm{H} z^{\frac{3}{2}} - \frac{8\xi_0 {\mathrm H}^3z^{\frac{5}{2}}}{(a_0\phi_0)^2} + \mathrm H (a_0\phi_0)^2 \sqrt z\end{split},\ee
where in the last expression, we have flipped $\phi$ to $z$, using the expression $\phi = {a_0\phi_0\over \sqrt z}$, to express $S_0$ in terms of $z$. It may be mentioned that equation (\ref{hbar0}) has never been solved for $S_0$, since it's extremely difficult, if not impossible. We have rather found $S_0$ using the classical solutions (\ref{11}). So, for consistency, one can trivially check that the expression for $S_0$ \eqref{S0} so obtained, satisfies equation (\ref{hbar0}) identically. In fact it should, because, equation (\ref{hbar0}) coincides with Hamiltonian constraint equation (\ref{3}) for $k=0$. Moreover, one can also compute the zeroth order on-shell action. Using classical solution (\ref{11}) one may express all the variables in terms of $t$ and substituting in the action (\ref{17}), one obtains

\be A_2 = \int \left[-\frac{3}{4\pi G} {a_0}^3 {\mathrm{H}}^2 e^{3\mathrm{H} t} - \frac{40\xi_0 {a_0}^3 {\mathrm{H}}^4}{{\phi_0}^2}e^{5\mathrm{H} t} + {a_0}^3 {\mathrm{H}}^2{\phi_0}^2e^{\mathrm{H} t}\right]dt.\ee
On integration, one thus finds
\be\begin{split} -\frac{1}{4\pi G} {a_0}^3 {\mathrm{H}} e^{3\mathrm{H} t} - \frac{8\xi_0 {a_0}^3 {\mathrm{H}}^3}{{\phi_0}^2}e^{5\mathrm{H} t} + {a_0}^3 {\mathrm{H}}{\phi_0}^2e^{\mathrm{H} t} = - \frac{1}{4\pi G}\mathrm{H} z^{\frac{3}{2}} - \frac{8\xi_0 {\mathrm H}^3z^{\frac{5}{2}}}{(a_0\phi_0)^2} + \mathrm H (a_0\phi_0)^2 \sqrt z,\end{split}\ee
which is the same expression (\ref{S0}). Therefore up-to zeroth order approximation, the wave function reads
\be \psi = \psi_0 e^{\frac{i}{\hbar}\left[- \frac{1}{4\pi G}\mathrm{H} z^{\frac{3}{2}} - \frac{8\xi_0 {\mathrm H}^3z^{\frac{5}{2}}}{(a_0\phi_0)^2} + \mathrm H (a_0\phi_0)^2 \sqrt z\right]}.\ee

\subsubsection{ First order approximation}
Now for $n=-1$, equation (\ref{hbar1}) can be expressed as,
\be -\frac{\sqrt z}{36\beta \xi x}\left(i S_{0,xx} - 2S_{0,x}S_{1,x} - \frac{i}{x}S_{0,x}\right) -\frac{1}{2 x z^{\frac{3}{2}}}\left(i S_{0,\phi\phi} - 4S_{0,\phi}S_{1,\phi}\right) + S_{1,z} + \frac{\xi_0x^2}{z^3\phi^3} S_{1,\phi} - \frac{3i \xi_0x^2}{z^3\phi^4} = 0. \ee
\noindent
Using the expression for $S_0$ obtained in (\ref{S0}), one can express $S_{1,z}$ in view of the above equation as
\be S_{1,z} = -\frac{i\left[ \frac{6\xi_0\mathrm H^2}{(a_0\phi_0)^4} z-\frac{3}{8}{1\over z} \right]}{1 - \frac{(a_0\phi_0)^3}{z^{3\over 2}} + {16\over 5}\frac{\xi_0\mathrm H^2}{a_0\phi_0}\sqrt z - \frac{4\xi_0\mathrm H^2}{(a_0\phi_0)^2}z}, \ee
In principle, one can integrate to obtain $S_1$ in the form,
\be S_1 = - i f(z).\ee
Therefore, the wavefunction to first-order approximation reads
\be \psi = \psi_{01} e^{\frac{i}{\hbar}\left[- \frac{\mathrm{H}}{4\pi G} z^{\frac{3}{2}} - \frac{8\xi_0 {\mathrm H}^3}{(a_0\phi_0)^2}z^{\frac{5}{2}} + \mathrm H (a_0\phi_0)^2 \sqrt z\right]},\ee
where,
\be \psi_{01} = \psi_0 e^{f(z)}.\ee
Thus, first-order approximation only modifies the prefactor, keeping the oscillatory behavior of the wave function intact. Since the wave function is oscillatory about the classical inflationary solution, so the correspondence between the quantum equation and the classical equations has been established.

\section{Concluding remarks}
Einstein-Gauss-Bonnet-Dilatonic coupled action successfully explains late-time accelerated expansion of the universe, fitting all the observed cosmological data fairly well. However, such a term appears under weak energy limit of heterotic string theory and so contributes at the early stage of cosmological evolution as well. Hence, it is important to study the quantum dynamics. Unfortunately, Hamiltonian structure of such an action does not exist, due to the problem of branching. We have associated an additional scalar curvature squared ($R^2$) term with dilatonic coupling in the action, to bypass the issue of branching. It is important to mention that alleviation of the issue of branching by the introduction of higher order curvature invariant term is not an artefact of minisuperspace approach, in which many degrees of freedom is suppressed. Note that $R^2$ itself contains terms with higher degree, e.g. $\dot a^4$, in Robertson-Walker minisuperspace. Canonical formulation of $\alpha R + \beta R^2$ action has been presented by Boulware \cite{Boul} in the whole superspace, which didn't encounter the issue of branching, even in the presence of momentum constraints. However, since Boulware's approach is limited for a particular form of action, it's not possible to follow the approach for the more general action under present consideration. To start with, we have considered the simplest homogeneous and isotropic minisuperspace model. A viable quantum version of action (\ref{A2}) has been presented, where effective Hamiltonian is hermitian and standard quantum mechanical probabilistic interpretation holds. Semiclassical approximation yields a wave function which has oscillatory behaviour about classical inflationary solution. Therefore, (\ref{A2}) appears to be a complete action which can successfully explain the history of cosmic evolution from the very early stage till date. Although, in view of our earlier discussion, we strongly believe that the presence of momentum constraints should not tell upon the results obtained, it is of-course important to check if it really does. However, it requires to consider a more general metric consisting of ($0,i$) components, where $i$ runs from $1$ to $3$. This largely complicates the Hamiltonian and presently appears extremely difficult to handle. This may be posed in future.

\appendix
\section{Matching the field equations}
It is now only left to prove that (\ref{3}) is the correct Hamiltonian description of the theory (\ref{A2}) under consideration. This is important, since an auxiliary variable $Q$ (\ref{Q1}) has been introduced to cast the action in canonical form, while no algebraic constraint has been added in the action to maintain the equivalence between the preceding (\ref{15}) and following actions (\ref{17}). However, it is enough to show that the Hamilton constraint equation (\ref{1}) is equivalent to the $(^0_0)$ component of Einstein's equation. This is because, replacing $p_x$ by $NQ$ and $x$ by $-{p_Q\over N}$, the phase-space description of the Hamilton constraint equation (\ref{3}) reduces to that presented in (\ref{2}). Further, using the definition of momenta (22), one gets back the Hamiltonian (\ref{1}).\\

\noindent
Before we proceed, let us remember that the $(^0_0)$ component of Einstein's equation, when multiplied by $\sqrt{-g}$, gives the Hamilton constraint equation. That is, if we start from the following Einstein-Hilbert action being minimally coupled to a scalar field
\be\label{AA1} A =\int\sqrt{-g}\;d^4x\left[\frac{R}{2\kappa} - \frac{1}{2} \phi_{,\mu}\phi^{,\mu}-V(\phi)\right] + \Sigma_R,\ee
then under metric variation, one obtains the field equation in the form
\be\label{E} \frac{G_{\mu\nu}}{\kappa} - T_{\mu\nu} = 0,\ee
where $G_{\mu\nu} = \left(R_{\mu\nu} - \frac{1}{2}g_{\mu\nu}R \right)$ and $T_{\mu\nu} = \left(\nabla_\mu\phi\nabla_\nu\phi - \frac{1}{2}g_{\mu\nu}\nabla_\lambda\phi\nabla^\lambda\phi - g_{\mu\nu}V \right)$ are the Einstein tensor the energy-momentum tensor respectively. The Hamiltonian is then

\be \label{Hc} H_c = \sqrt{-g}\left[{G^0_0\over \kappa} - T^0_0\right] = 0.\ee
This may be proved following standard ADM (Arnwitt, Deser and Misner) formalism \cite{ADM}. The ADM Hamiltonian is
\be\label{AHc} H_{c} = N\left[ \frac{2\kappa}{\sqrt h}\left(p^{ij}p_{ij} - \frac{1}{2}p^2\right) - \frac{\sqrt h~^{(3)}R}{2\kappa} + \frac{p_\phi^2}{\sqrt h} + \sqrt h V\right] -2N_i{p^{ij}}_{||j} = N\mathcal{H} + N_i \mathcal{H}^i,\ee
where, $ p^{ij} = -\frac{\sqrt h}{2\kappa}\left( K^{ij} - Kh^{ij} \right)$, $p = \frac{\sqrt h K}{\kappa}$ and $p_\phi = \frac{\sqrt h}{N} \phi_{,0}$. $K^{ij}$ and $h^{ij}$ are the extrinsic curvature tensor and the three metric respectively, while, $K$ and $h$ are the corresponding traces. $N$ is the lapse function and $N_i$ is the shift vector. Further, ${p^{ij}}_{||j}$ is the covariant derivative of $p^{ij}$ in three space. Now in the background of Robertson-Walker metric (\ref{rw}), both the above forms (\ref{Hc}) and (\ref{AHc}) of Hamiltonian lead to
\be\label{AHc1} H_c = - \frac{3a}{\kappa}\left(\frac{\dot a^2}{N} + k N\right) + a^3\left(\frac{\dot\phi^2}{2N} + V N\right) = 0.\ee
With the above understanding, let us start with the action (\ref{A2}), which under metric variation leads to the following field equation \cite{field, field1}
\be\label{gfe}\begin{split}
&\frac{G_{\mu\nu}}{\kappa} - T_{\mu\nu} + 2\beta\left( 2\xi(\phi)RR_{\mu\nu} + 2g_{\mu\nu}\Box( \xi(\phi)R) - 2\nabla_\mu\nabla_\nu( \xi(\phi)R) - \frac{1}{2}g_{\mu\nu}\xi(\phi)R^2 \right) + 2\xi(\phi) H_{\mu\nu} \\
&+ 8\left({\xi'}^2\nabla^\rho\phi\nabla^\sigma\phi + \xi'\nabla^\rho\nabla^\sigma\phi \right)P_{\mu\rho\nu\sigma} = 0
\end{split}\ee
where
\begin{subequations}\begin{align}
H_{\mu\nu} &= 2\left(RR_{\mu\nu} - 2R_{\mu\rho}R^\rho_\nu - 2R_{\mu\rho\nu\sigma}R^{\rho\sigma} + R_{\mu\rho\sigma\lambda}R_\nu^{\sigma\rho\lambda} \right) - \frac{1}{2}g_{\mu\nu}\mathcal{G} \\
P_{\mu\nu\rho\sigma} &= R_{\mu\nu\rho\sigma} + 2g_{\mu[\sigma}R_{\rho]\nu} + 2g_{\nu[\rho}R_{\sigma]\mu} + Rg_{\mu[\rho}g_{\sigma]\nu}
\end{align}\end{subequations}
Now, in the background of Robertson-Walker metric (\ref{rw}), the ($^0_0$) component of the field equation (\ref{gfe}) takes the following form
\be\label{00g}\begin{split} & -\frac{36\beta}{a^2N^4}\left(2\dot a~\dddot a - 2\dot a^2 \frac{\ddot N}{N} - \ddot a^2 - 4\dot a \ddot a \frac{\dot N}{N} + 2 \dot a^2 \frac{\ddot a}{a} + 5 \dot a \frac{\dot N^2}{N^2} - 2 \frac{\dot a^3\dot N}{aN} - 3\frac{\dot a^4}{a^2} - 2k N^2 \frac{\dot a^2}{a^2} + \frac{k^2N^4}{a^2}\right)\\
& -\frac{72\beta \xi'\dot a\dot\phi}{a^3N^4}\left(a\ddot a + \dot a^2 + k N^2 - \frac{a\dot a \dot N}{N}\right) - \frac{3}{\kappa a^2}\left(\frac{\dot a^2}{N^2} + k\right) + \left(\frac{\dot\phi^2}{2N^2} + V\right) - \frac{24\xi'\dot\phi\dot a}{N^2a^3}\left(\frac{\dot a^2}{N^2} + k\right) = 0.\end{split}\ee
The Hamiltonian $H_c$ (\ref{1}) on the other hand, may be expressed in terms of the scale factor $a$ as
\be\label{Hg}\begin{split} & H_c = -\frac{36\beta a}{N^3}\left(2\dot a~\dddot a - 2\dot a^2 \frac{\ddot N}{N} - \ddot a^2 - 4\dot a \ddot a \frac{\dot N}{N} + 2 \dot a^2 \frac{\ddot a}{a} + 5 \dot a \frac{\dot N^2}{N^2} - 2 \frac{\dot a^3\dot N}{aN} - 3\frac{\dot a^4}{a^2} - 2k N^2 \frac{\dot a^2}{a^2} + \frac{k^2N^4}{a^2}\right)\\
& -\frac{72\beta \xi'\dot a\dot\phi}{N^3}\left(a\ddot a + \dot a^2 + k N^2 - \frac{a\dot a \dot N}{N}\right) - \frac{3a}{\kappa}\left(\frac{\dot a^2}{N} + k N\right) + a^3\left(\frac{\dot\phi^2}{2N} + V N\right) - \frac{24\xi'\dot\phi\dot a}{N}\left(\frac{\dot a^2}{N^2} + k\right) = 0.\end{split}\ee
Clearly, multiplying equation (\ref{00g}) by the same factor $\sqrt{-g} = N a^3$, the Hamiltonian (\ref{Hg}) is realized. Thus, the Hamiltonian obtained in the present methodology is the Hamiltonian of the action (\ref{A2}) under consideration. Now substituting $\mathcal{H}$ from equation (\ref{3}), $p_z$ from equation (\ref{12}), $p_\phi$ from equation (\ref{13}), $p_x = NQ$, $x=\frac{\dot z}{N}$ and $Q$ from equation (\ref{Q1}), the canonical action (\ref{CA}) reduces to the action (\ref{15}), which is essentially action (\ref{A2}) in Robertson-Walker (\ref{rw}) minisuperspace.

\end{document}